\documentclass[12pt,reqno]{amsart}
\usepackage{graphicx}
\usepackage{amscd}
\usepackage{amsmath}
\usepackage{epsfig}
\usepackage{amsfonts}
\usepackage{amssymb}

\providecommand{\U}[1]{\protect\rule{.1in}{.1in}}
\providecommand{\U}[1]{\protect\rule{.1in}{.1in}}
\allowdisplaybreaks

\theoremstyle{plain}

\numberwithin{equation}{section}

\input{tcilatex}

\begin{document}
\title[A Graphical Approach to a Model of Dendritic Tree]{A Graphical
Approach to a Model of Neuronal Tree with Variable Diameter}
\author{Marco Herrera-Vald\'{e}z }
\address{Mathematical, Computational and Modeling Sciences Center, Arizona
State University, Tempe, AZ 85287--1904, U.S.A.}
\email{curandero13@gmail.com}
\author{Sergei K. Suslov}
\address{Mathematical, Computational and Modeling Sciences Center, Arizona
State University, Tempe, AZ 85287--1904, U.S.A.}
\email{sks@asu.edu}
\urladdr{http://hahn.la.asu.edu/\symbol{126}suslov/index.html}
\date{\today }

\begin{abstract}
We propose a simple graphical approach to steady state solutions of the
cable equation for a general model of dendritic tree with tapering. A simple
case of transient solutions is also briefly discussed.
\end{abstract}

\maketitle

\section{Introduction}

The function of many physiological systems depends on branched structures
that exist both at the tissue (e.g. nervous plexi, lungs, and the vascular
and lymphatic systems) and the cellular level (e.g. neurons). Of particular
interest, local and global propagation of electrical signals within the
nervous system depends on the integration, processing, and further
generation of electrical pulses that travel through neurons. In turn, the
tree-like morphology of neurons facilitates simultaneous signaling to cells
located in different places and over long distances.

Neuronal morphology is typically modeled by assuming that the shape of small
neuronal segments, or \textit{neurites}, is approximated by cylinders of
different diameters. As a consequence, cable theory \cite{Rall59}, \cite%
{Rall60}, \cite{Rall62}, \cite{Rall69}, \cite{Rall77}, \cite{Rall89} plays a
central role in the theoretical and experimental study of electrical
conduction in neurons; see, for example \cite{Baer:Tier86}, \cite%
{Bluman:Tuckwell87}, \cite{Tuckwell88}, \cite{Coombesetal07}, \cite%
{Cox:Raol04}, \cite{Durand84}, \cite{Evans05}, \cite{Evans:Kember:Major92},
\cite{Evans:Major:Kember95}, \cite{Jack:Noble:Tsien}, \cite{Schier:Ohme},
\cite{Surkisetal96} and references therein. Notably, one of the most
interesting results from recent theoretical work is that geometrical
properties of neuronal membranes may exert powerful effects on signal
propagation even in the presence of voltage-dependent channels \cite%
{Vett:Roth:Haus01}.

Theoretical research involving realistic neuronal morphologies is typically
done by numerically solving systems of cable equations defined on cylinders
with different radii, and assuming that voltage and current are continuous
functions of space and time. To the best of our knowledge, graphical methods
seem have not been widely applied yet in the mathematical modeling of
neurons. Graphical methods are very useful and popular in different branches
of modern physics. It is worth noting, for example, Feynman diagrams in
quantum mechanical or statistical field theory \cite{Abrik:Gork:Dzyal}, \cite%
{Akh:Ber}, \cite{Ber:Lif:Pit}, \cite{Feynman49a}, \cite{Feynman49b}, \cite%
{FynmanQED}, \cite{Kaiser05}, \cite{Mattuck}, \cite{Wein},
Vilenkin-Kuznetsov-Smorodinskii approach to solutions of $n$-dimensional
Laplace equation \cite{Me:Co:Su}, \cite{Ni:Su:Uv}, \cite{Smir:Shit}, \cite%
{Smor76}, applications in solid-state theory, etc. A goal of this paper is
to make a modest step in this direction (see also \cite{Abbott92}, \cite%
{Coombesetal07} and references therein). We use explicit solutions from
recent papers on variable quadratic Hamiltonians in nonrelativistic quantum
mechanics \cite{Chru:Jurk}, \cite{Cor-Sot:Lop:Sua:Sus}, \cite%
{Cor-Sot:Sua:Sus}, \cite{Cor-Sot:Sua:SusInv}, \cite{Cor-Sot:Sus}, \cite%
{Lan:Sus}, \cite{Me:Co:Su}, \cite{Sua:Sus:Vega}, \cite{Suslov10} to describe
steady state and transient solutions to linear cable equations modeling
neurites with non-necessarily constant radius.

\section{Cable Equation with Varying Radius}

At a closer view, neurites can be regarded as volumes of revolution, defined
by rotating a smooth function $r=r(x)$ representing the local radius of the
neurite where $x$ represents distance along the neurite. As a result, the
cable theory implies the following set of equations \cite{Jack:Noble:Tsien},
\cite{Rall62}:%
\begin{equation}
2\pi rI_{m}\frac{ds}{dx}=-\frac{\partial I}{\partial x},\qquad \frac{ds}{dx}=%
\sqrt{1+\left( \frac{dr}{dx}\right) ^{2}}  \label{cable1}
\end{equation}%
\begin{equation}
I_{m}=\frac{V}{R_{m}}+C_{m}\frac{\partial V}{\partial t},  \label{cable2}
\end{equation}%
\begin{equation}
I=-\frac{\pi r^{2}}{R_{i}}\frac{\partial V}{\partial x}.  \label{cable3}
\end{equation}%
Here, $V$ represents the voltage difference across the membrane (interior
minus exterior) as a deviation from its resting value, $I_{m}$ is the
membrane current density, $I=I_{a}$ is the total axial current, $R_{m}$ is
the membrane resistance, $R_{i}$ is the intercellular resistivity and $C_{m}$
is membrane capacitance (more details can be found in \cite{Jack:Noble:Tsien}%
, \cite{Rall59} and \cite{Rall62}). Differentiating equation~(\ref{cable3})
with respect to $x$ and substituting the result into (\ref{cable1}) with the
help of (\ref{cable2}) one gets%
\begin{equation}
\left( V+C_{m}R_{m}\frac{\partial V}{\partial t}\right) \sqrt{1+\left( \frac{%
dr}{dx}\right) ^{2}}=\frac{R_{m}}{2R_{i}}\frac{1}{r}\frac{\partial }{%
\partial x}\left( r^{2}\frac{\partial V}{\partial x}\right) ,
\label{CableEquation}
\end{equation}%
which is the cable equation with tapering for a single branch of dendritic
tree (see \cite{Cox:Raol04}, \cite{Jack:Noble:Tsien}, \cite{Rall62} and \cite%
{Surkisetal96} for more details).

We shall be particularly interested in solutions of the cable equation (\ref%
{CableEquation}) corresponding to termination with a \textquotedblleft
sealed end\textquotedblright , namely, when at the end point $x=x_{1}$ the
membrane cylinder is sealed with a disk composed of the same membrane. In
this case, the corresponding boundary condition can be derived by setting%
\begin{equation}
I_{a}=\pi r^{2}I_{m},  \label{sealingend}
\end{equation}%
at $x=x_{1}.$ Then, in view of (\ref{cable2})--(\ref{cable3}), one gets \cite%
{Rall59}:%
\begin{equation}
\left. \left( V+C_{m}R_{m}\frac{\partial V}{\partial t}+\frac{R_{m}}{R_{i}}%
\frac{\partial V}{\partial x}\right) \right\vert _{x=x_{1}}=0,\qquad t\geq 0.
\label{sealedendboundary}
\end{equation}%
In a similar fashion, at the somatic end one gets%
\begin{equation}
\left. \left( V+C_{s}R_{s}\frac{\partial V}{\partial t}-\frac{R_{s}}{R_{i}}%
\frac{\partial V}{\partial x}\right) \right\vert _{x=x_{0}}=0,\qquad t\geq 0,
\label{somaendboundary}
\end{equation}%
where $R_{s}$ is the somatic resistance and $C_{s}$ is the somatic
capacitance \cite{Durand84}. We shall use these conditions for the
steady-state and transient solutions of the cable equation. (Later we may
impose similar boundary conditions at the points of branching.)

In this Letter, we shall first concentrate on steady-state solutions of the
cable equation, when $\partial V/\partial t\equiv 0.$ Then%
\begin{equation}
V\sqrt{1+\left( \frac{dr}{dx}\right) ^{2}}=\frac{R_{m}}{2rR_{i}}\frac{d}{dx}%
\left( r^{2}\frac{dV}{dx}\right) \qquad \left( x_{0}\leq x\leq x_{1}\right)
\label{Sturm}
\end{equation}%
and%
\begin{equation}
\left. V\right\vert _{x=x_{0}}=V_{0},\qquad \left. \left( \frac{dV}{dx}%
+B\left( x\right) V\right) \right\vert _{x=x_{1}}=0,\qquad B\left(
x_{1}\right) =\frac{R_{i}}{R_{m}}.  \label{SturmBoundary}
\end{equation}%
This boundary value problem can be conveniently solved (by a direct
substitution for each branch of the dendritic tree) in terms of standard
solutions of this second order ordinary differential equation as follows%
\begin{equation}
V\left( x\right) =V\left( x_{0}\right) \frac{C\left( x\right) +B\left(
x_{1}\right) S\left( x\right) }{C\left( x_{0}\right) +B\left( x_{1}\right)
S\left( x_{0}\right) },  \label{Solution}
\end{equation}%
where $C\left( x\right) $ and $S\left( x\right) $ are two linearly
independent solutions of the stationary cable equation (\ref{Sturm}) that
satisfy special boundary conditions $C\left( x_{1}\right) =1,$ $C^{\prime
}\left( x_{1}\right) =0$ and $S\left( x_{1}\right) =0,$ $S^{\prime }\left(
x_{1}\right) =-1.$ Then%
\begin{equation}
\frac{dV}{dx}+B\left( x\right) V=0\qquad \left( x_{0}\leq x\leq x_{1}\right)
\label{Boundary}
\end{equation}%
with the corresponding current density/voltage ratio function $B\left(
x\right) $ given by%
\begin{equation}
B\left( x\right) =-\frac{V^{\prime }}{V}=-\frac{C^{\prime }\left( x\right)
+B\left( x_{1}\right) S^{\prime }\left( x\right) }{C\left( x\right) +B\left(
x_{1}\right) S\left( x\right) }  \label{Ratio}
\end{equation}%
in term of the standard solutions $C\left( x\right) $ and $S\left( x\right)
. $ Throughout this Letter, we shall refer to a case, when $B\left( x\right)
>0\ \left( x_{0}<x<x_{1}\right) ,$ as the case of weak tapering. An opposite
situation, when $B\left( \xi \right) =0$ at certain point $x_{0}<\xi <x_{1}$
and an inverse of the current may occur, shall be called a case of the
strong tapering. (A case of strong tapering has been numerically discovered
in \cite{MTBI}.)

\section{Tapering with Analytic, Asymptotic and/or Numerical Solutions}

In this Letter, we consider a general model of a dendrite as a (binary)
directed tree (from the soma to its terminal ends) consisting of axially
symmetric branches with the following types of tapering.

\subsection{Cylinder}

Here, $r=r_{0}=$constant, $0\leq x\leq L$ and the cable equation takes the
simplest form%
\begin{equation}
\lambda ^{2}\frac{d^{2}V}{dx^{2}}=V,\qquad \lambda ^{2}=\frac{r_{0}R_{m}}{%
2R_{i}}  \label{CableCylinder}
\end{equation}%
with a familiar solution \cite{Jack:Noble:Tsien}, \cite{Rall59}, \cite%
{Rall62}:%
\begin{equation}
V\left( x\right) =V_{0}\frac{\cosh \left( \left( L-x\right) /\lambda \right)
+\lambda B_{L}\sinh \left( \left( L-x\right) /\lambda \right) }{\cosh \left(
L/\lambda \right) +\lambda B_{L}\sinh \left( L/\lambda \right) }
\label{CableCylinderSolution}
\end{equation}%
subject to boundary conditions%
\begin{equation}
V\left( 0\right) =V_{0},\qquad B_{L}V\left( L\right) +\frac{dV}{dx}\left(
L\right) =0.  \label{CableConeBoundary}
\end{equation}%
Then%
\begin{equation}
B\left( x\right) =\frac{\lambda B_{L}+\tanh \left( \left( L-x\right)
/\lambda \right) }{\lambda +\lambda ^{2}B_{L}\tanh \left( \left( L-x\right)
/\lambda \right) }.  \label{CylinderB}
\end{equation}%
(See \cite{Jack:Noble:Tsien}, \cite{Rall62}, \cite{Rall77}, \cite{Rall89},
\cite{Tuckwell88} and references therein for more details.)

\subsection{Frustum(Cone)}

Here, $r=r\left( x\right) =r_{0}+\frac{r_{1}-r_{0}}{L}x$ with $0\leq x\leq
L. $ The steady-state solution of the corresponding cable equation%
\begin{equation}
r\frac{d^{2}V}{dx^{2}}+2\frac{dV}{dx}=\frac{\mu ^{2}}{4}V,\qquad \mu =\sqrt{%
\frac{8R_{i}}{R_{m}}}\frac{\left( 1+\left( \frac{r_{1}-r_{0}}{L}\right)
^{2}\right) ^{1/4}}{\left\vert \frac{r_{1}-r_{0}}{L}\right\vert }
\label{CableCone}
\end{equation}%
subject to boundary conditions (\ref{CableConeBoundary}) are given by \cite%
{Baer:Herr:Sus:Vega}%
\begin{equation}
V\left( x\right) =V_{0}\frac{C\left( L-x\right) +B_{L}S\left( L-x\right) }{%
C\left( L\right) +B_{L}S\left( L\right) }.  \label{CableConeSolution}
\end{equation}%
Here, the standard solutions that satisfy $S\left( 0\right) =0,$ $S^{\prime
}\left( 0\right) =1$ and $C\left( 0\right) =1,$ $C^{\prime }\left( 0\right)
=0$ can be constructed as follows%
\begin{equation}
S\left( x\right) =\frac{2Lr_{0}^{3/2}}{\left( r_{1}-r_{0}\right) \sqrt{r}}%
\left[ K_{1}\left( \mu \sqrt{r_{0}}\right) I_{1}\left( \mu \sqrt{r}\right)
-I_{1}\left( \mu \sqrt{r_{0}}\right) K_{1}\left( \mu \sqrt{r}\right) \right]
\label{SBessel}
\end{equation}%
and%
\begin{eqnarray}
C\left( x\right) &=&\left[ \mu \sqrt{r_{0}}K_{0}\left( \mu \sqrt{r_{0}}%
\right) +2K_{1}\left( \mu \sqrt{r_{0}}\right) \right] \sqrt{\frac{r_{0}}{%
r_{1}}}I_{1}\left( \mu \sqrt{r}\right)  \label{CBessel} \\
&&+\left[ \mu \sqrt{r_{0}}I_{0}\left( \mu \sqrt{r_{0}}\right) -2I_{1}\left(
\mu \sqrt{r_{0}}\right) \right] \sqrt{\frac{r_{0}}{r_{1}}}K_{1}\left( \mu
\sqrt{r}\right)  \notag
\end{eqnarray}%
in terms of modified Bessel functions $I_{\nu }\left( z\right) $ and $K_{\nu
}\left( z\right) $ of orders $\nu =0,1$ (different aspects of the advanced
theory of Bessel functions can be found in \cite{Ab:St}, \cite{An:As:Ro},
\cite{Askey}, \cite{Erd}, \cite{Ni:Uv}, \cite{Olver}, \cite{Rain}, \cite{Vil}
and \cite{Wa}).

\subsection{Hyperbola}

If $r=a\cosh \left( \frac{x-b}{a}\right) $ on an interval $x_{0}\leq x\leq
x_{1},$ the cable equation (\ref{CableEquation}) takes the form%
\begin{equation}
\left( V+C_{m}R_{m}\frac{\partial V}{\partial t}\right) \cosh \left( \frac{%
x-b}{a}\right) =\frac{R_{m}}{2R_{i}}\left[ 2\sinh \left( \frac{x-b}{a}%
\right) \frac{\partial V}{\partial x}+a\cosh \left( \frac{x-b}{a}\right)
\frac{\partial ^{2}V}{\partial x^{2}}\right] .  \label{hypercable}
\end{equation}%
This special case of tapering is integrable in terms of elementary functions
\cite{Herr:Sus} (see also \cite{Chru:Jurk} and \cite{Cor-Sot:Sua:SusInv} for
a similar problem related to a model of the dumped quantum oscillator). For
the steady-state solutions one obtains the following equation%
\begin{equation}
V^{\prime \prime }+2\lambda \tanh \left( \lambda x+\delta \right) V^{\prime
}=\mu _{0}^{2}V  \label{hypersteady}
\end{equation}%
with new parameters%
\begin{equation}
\lambda =\frac{1}{a},\qquad \delta =-\lambda b,\qquad \mu _{0}^{2}=\frac{%
2R_{i}}{aR_{m}}.  \label{hypernotations}
\end{equation}%
The corresponding two linearly independent solutions, namely,%
\begin{equation}
V_{1}\left( x\right) =\frac{\sinh \left( \mu x+\gamma \right) }{\cosh \left(
\lambda x+\delta \right) },\qquad V_{2}\left( x\right) =\frac{\cosh \left(
\mu x+\gamma \right) }{\cosh \left( \lambda x+\delta \right) },\qquad \mu =%
\sqrt{\mu _{0}^{2}+\lambda ^{2}},  \label{hyperlinsols}
\end{equation}%
can be verified by a direct substitution for an arbitrary parameter $\gamma
. $

The required steady-state solution of the boundary value problem%
\begin{equation}
V\left( x_{0}\right) =V_{0},\qquad BV\left( x_{1}\right) +\frac{dV}{dx}%
\left( x_{1}\right) =0  \label{hypercableboundary}
\end{equation}%
is given by%
\begin{equation}
V\left( x\right) =V_{0}\frac{C\left( x_{1}-x\right) +BS\left( x_{1}-x\right)
}{C\left( x_{1}-x_{0}\right) +BS\left( x_{1}-x_{0}\right) },
\label{hypercablesolution}
\end{equation}%
where%
\begin{eqnarray}
&&C\left( x_{1}-x\right) =\cosh \left( \lambda \left( b-x_{1}\right) \right)
\frac{\cosh \left( \mu \left( x_{1}-x\right) \right) }{\cosh \left( \lambda
\left( x-b\right) \right) }  \label{hypercableC} \\
&&\quad \qquad \qquad \quad +\sinh \left( \lambda \left( b-x_{1}\right)
\right) \frac{\lambda \sinh \left( \mu \left( x_{1}-x\right) \right) }{\mu
\cosh \left( \lambda \left( x-b\right) \right) }  \notag
\end{eqnarray}%
and%
\begin{equation}
S\left( x_{1}-x\right) =\cosh \left( \lambda \left( b-x_{1}\right) \right)
\frac{\sinh \left( \mu \left( x_{1}-x\right) \right) }{\mu \cosh \left(
\lambda \left( x-b\right) \right) }.  \label{hypercableS}
\end{equation}%
(See \cite{Herr:Sus} for more details.)

\subsection{General Case of Axial Symmetry}

One can use numerical methods and/or WKB-type approximation in order to
obtain standard solutions. For example \cite{Ni:Uv},%
\begin{equation}
V\approx \frac{1}{\sqrt{r^{2}\left( x\right) p\left( x\right) }}\left[
Ae^{\xi \left( x\right) }+Be^{-\xi \left( x\right) }\right] ,  \label{WKB}
\end{equation}%
where%
\begin{equation}
p\left( x\right) =\left( \frac{2}{r}\frac{ds}{dx}\frac{R_{i}}{R_{m}}\right)
^{1/2},\qquad \xi \left( x\right) =\int_{x_{0}}^{x}p\left( t\right) \ dt.
\label{WKBP}
\end{equation}%
(See \cite{Jack:Noble:Tsien} and \cite{Rall62} for further details.)

\section{A Graphical Approach}

Graphical rules for steady-state voltages and currents in a model of
dendritic tree with tapering are as follows.

\subsection{ Single Axially Symmetric Branch with Arbitrary Tapering}

For a single branch with tapering voltage and current density/voltage ratio
are given by%
\begin{eqnarray}
V\left( x\right) &=&\frac{C\left( x_{1}-x\right) +B\left( x_{1}\right)
S\left( x_{1}-x\right) }{C\left( x_{1}-x_{0}\right) +B\left( x_{1}\right)
S\left( x_{1}-x_{0}\right) }V\left( x_{0}\right)  \label{RuleV} \\
&=&\left( C\left( x_{1}-x\right) +B\left( x_{1}\right) S\left(
x_{1}-x\right) \right) V\left( x_{1}\right) ,  \notag
\end{eqnarray}%
\begin{equation}
B\left( x\right) =\frac{C^{\prime }\left( x_{1}-x\right) +B\left(
x_{1}\right) S^{\prime }\left( x_{1}-x\right) }{C\left( x_{1}-x\right)
+B\left( x_{1}\right) S\left( x_{1}-x\right) }=-\frac{V^{\prime }\left(
x\right) }{V\left( x\right) },  \label{RuleB}
\end{equation}%
respectively (see Figure~1).

%
%Figure1%
\begin{figure}[htbp]
\centering \scalebox{.75}{\includegraphics{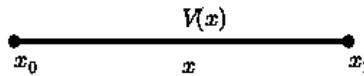}}
\caption{A Single Branch with Tapering.}
\end{figure}

\subsection{Junction of Three Branches with Different Types of Tapering}

The internal potential and current are assumed to be continuous at all
dendritic branch points and at the soma-dendritic junction \cite{Rall59}. We
consider a general case when each branch has its own tapering, say $%
r=r\left( x\right) ,$ $r_{1}=r_{1}\left( x\right) $ and $r_{2}=r_{2}\left(
x\right) $ (see Figure~2). Then%
\begin{equation}
V\left( x_{12}\right) =V_{1}\left( x_{12}\right) =V_{2}\left( x_{12}\right) ,
\label{RulesVVV}
\end{equation}%
\begin{equation}
r^{2}\left( x_{12}\right) B\left( x_{12}\right) =r_{1}^{2}\left(
x_{12}\right) B_{1}\left( x_{12}\right) +r_{2}^{2}\left( x_{12}\right)
B_{2}\left( x_{12}\right) .  \label{RulesBBB}
\end{equation}

%
%Figure2%
\begin{figure}[htbp]
\centering \scalebox{.75}{\includegraphics{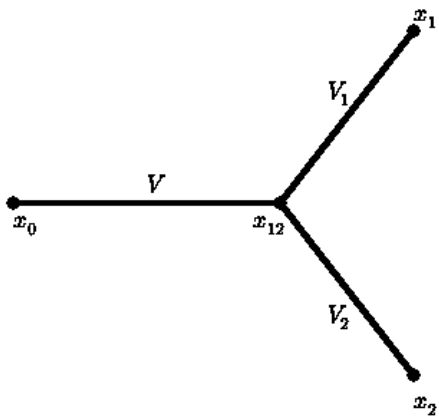}}
\caption{A Junction of Three Different Branches.}
\end{figure}

The total ratio constant $B\left( x_{12}\right) $ at the branching point $%
x_{12}$ is given by the following expression%
\begin{eqnarray}
B\left( x_{12}\right) &=&\frac{r_{1}^{2}\left( x_{12}\right) }{r^{2}\left(
x_{12}\right) }B_{1}\left( x_{12}\right) +\frac{r_{2}^{2}\left(
x_{12}\right) }{r^{2}\left( x_{12}\right) }B_{2}\left( x_{12}\right)
\label{BranchingB} \\
&=&B\left( B_{1}\left( x_{1}\right) ,B_{2}\left( x_{2}\right) \right)  \notag
\\
&=&\frac{r_{1}^{2}\left( x_{12}\right) }{r^{2}\left( x_{12}\right) }\frac{%
C_{1}^{\prime }\left( x_{1}-x_{12}\right) +B_{1}\left( x_{1}\right)
S_{1}^{\prime }\left( x_{1}-x_{12}\right) }{C_{1}\left( x_{1}-x_{12}\right)
+B_{1}\left( x_{1}\right) S_{1}\left( x_{1}-x_{12}\right) }  \notag \\
&&+\frac{r_{2}^{2}\left( x_{12}\right) }{r^{2}\left( x_{12}\right) }\frac{%
C_{2}^{\prime }\left( x_{2}-x_{12}\right) +B_{2}\left( x_{2}\right)
S_{2}^{\prime }\left( x_{2}-x_{12}\right) }{C_{2}\left( x_{2}-x_{12}\right)
+B_{2}\left( x_{2}\right) S_{2}\left( x_{2}-x_{12}\right) }.  \notag
\end{eqnarray}%
Then the ratio constant $B\left( x_{0}\right) $ is%
\begin{equation}
B\left( x_{0}\right) =\frac{C^{\prime }\left( x_{12}-x_{0}\right) +B\left(
x_{12}\right) S^{\prime }\left( x_{12}-x_{0}\right) }{C\left(
x_{12}-x_{0}\right) +B\left( x_{12}\right) S\left( x_{12}-x_{0}\right) }
\label{BranchingB0}
\end{equation}%
with the coefficient $B\left( x_{12}\right) $ found by the previous formula (%
\ref{BranchingB}).

\subsection{Junction of $\left( n+1\right) $-Branches}

In a similar fashion, at the branching point $x_{\alpha },$ one gets%
\begin{equation}
V\left( x_{\alpha }\right) =V_{1}\left( x_{\alpha }\right) =V_{2}\left(
x_{\alpha }\right) =...=V_{n}\left( x_{\alpha }\right) ,  \label{RulesnV}
\end{equation}%
\begin{eqnarray}
r^{2}\left( x_{\alpha }\right) B\left( x_{\alpha }\right) &=&r_{1}^{2}\left(
x_{\alpha }\right) B_{1}\left( x_{\alpha }\right) +r_{2}^{2}\left( x_{\alpha
}\right) B_{2}\left( x_{\alpha }\right) +...+r_{n}^{2}\left( x_{\alpha
}\right) B_{n}\left( x_{\alpha }\right)  \label{RulesnB} \\
&=&\sum_{i=1}^{n}r_{i}^{2}\left( x_{\alpha }\right) B_{i}\left( x_{\alpha
}\right) .  \notag
\end{eqnarray}%
(See Figure~3.)

%
%Figure3%
\begin{figure}[htbp]
\centering \scalebox{.75}{\includegraphics{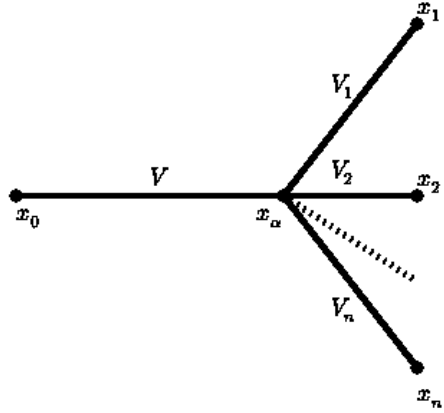}}
\caption{A Junction of $(n+1)$-Branches.}
\end{figure}

Combination of the above graphical rules results in a simple algorithm of
evaluation of voltages and currents in the model of dendritic tree under
consideration as follows.

Evaluate constants $B\left( x_{\alpha }\right) $ for all branching points of
the tree: (a) first apply formula (\ref{BranchingB}) for all open notes; (b)
remove the above nodes from the tree and keep repeating the previous step
until you reach the root of tree (soma).

In order to find voltage at a point $x$ of the dendritic tree, follow the
path $x_{0}\rightarrow x$ and multiply the initial voltage $V\left(
x_{0}\right) $ by consecutive corresponding factors from formula (\ref{RuleV}%
) changing at each intersection of the tree. The ratio of voltages $V\left(
x_{\alpha }\right) $ and $V\left( x_{\beta }\right) $ at two terminal
points, can be determine in a graphic form by the previous rule applied to
the shortest path $x_{\beta }\rightarrow x_{\beta }.$

\section{Examples}

Our formulas (\ref{BranchingB})--(\ref{BranchingB0}) define ratio
coefficients $B\left( x_{\alpha }\right) $ for all vertexes for the standard
node on Figure~2. For the corresponding voltages, one can write%
\begin{eqnarray}
V\left( x_{0}\right) &=&\left[ C\left( x_{12}-x_{0}\right) +B\left(
x_{12}\right) S\left( x_{12}-x_{0}\right) \right] V\left( x_{12}\right)
\label{Ex1} \\
&=&\left[ C\left( x_{12}-x_{0}\right) +B\left( x_{12}\right) S\left(
x_{12}-x_{0}\right) \right] V\left( x_{12}\right)  \notag \\
&&\times \left[ C\left( x_{1}-x_{12}\right) +B\left( x_{1}\right) S\left(
x_{1}-x_{12}\right) \right] V\left( x_{1}\right)  \notag \\
&=&\left[ C\left( x_{12}-x_{0}\right) +B\left( x_{12}\right) S\left(
x_{12}-x_{0}\right) \right] V\left( x_{12}\right)  \notag \\
&&\times \left[ C\left( x_{2}-x_{12}\right) +B\left( x_{2}\right) S\left(
x_{2}-x_{12}\right) \right] V\left( x_{2}\right)  \notag
\end{eqnarray}%
and%
\begin{equation}
\frac{V\left( x_{1}\right) }{V\left( x_{2}\right) }=\frac{C\left(
x_{2}-x_{12}\right) +B\left( x_{2}\right) S\left( x_{2}-x_{12}\right) }{%
C\left( x_{1}-x_{12}\right) +B\left( x_{1}\right) S\left(
x_{1}-x_{12}\right) }.  \label{Ex2}
\end{equation}%
Further examples are left to the reader.

\section{Transient Solutions}

\subsection{A Single Branch with Smooth Tapering}

Let us consider the cable equation (\ref{CableEquation}) for a single branch
with an arbitrary smooth tapering $r=r\left( x\right) $ on the interval $%
x_{0}\leq x\leq x_{1}.$ The separation of variables%
\begin{equation}
V\left( x,t\right) =e^{-\left( 1+\alpha ^{2}\right) t/\tau _{m}}U\left(
x\right) ,\qquad \tau _{m}=C_{m}R_{m}  \label{SepVar}
\end{equation}%
in results in%
\begin{equation}
\frac{1}{r}\frac{d}{dx}\left( r^{2}\frac{dU}{dx}\right) +\omega ^{2}\frac{ds%
}{dx}U=0,\qquad \omega ^{2}=\frac{2R_{i}}{R_{m}}\alpha ^{2},
\label{SepCableEq}
\end{equation}%
where $\alpha $ is a separation constant. The boundary condition at the
sealed end (\ref{sealedendboundary}) takes the form%
\begin{equation}
\left. \left( \frac{dU}{dx}-\frac{1}{2}\omega ^{2}U\right) \right\vert
_{x=x_{1}}=0.  \label{SepSealBoundary}
\end{equation}%
A general solution of this problem can be conveniently written (for each
branch of the dendritic tree) as follows%
\begin{equation}
U\left( x\right) =U\left( x,\omega \right) =A\left[ C\left( x,\omega \right)
+\frac{1}{2}\omega ^{2}S\left( x,\omega \right) \right] ,  \label{SepGenSol}
\end{equation}%
where $A$ is a constant and $C\left( x,\omega \right) $ and $S\left(
x,\omega \right) $ are two linearly independent standard solutions of
equation (\ref{SepCableEq}) that satisfy special boundary conditions $%
C\left( x_{1},\omega \right) =1,$ $C^{\prime }\left( x_{1},\omega \right) =0$
and $S\left( x_{1},\omega \right) =0,$ $S^{\prime }\left( x_{1},\omega
\right) =1.$ Then the boundary condition (\ref{somaendboundary}) at the
somatic end $x=x_{0},$ namely,%
\begin{equation}
\left. \left( \frac{dU}{dx}+\left[ \frac{R_{i}}{R_{s}}\left( \frac{\tau _{s}%
}{\tau _{m}}-1\right) +\frac{C_{s}}{2C_{m}}\omega ^{2}\right] U\right)
\right\vert _{x=x_{0}}=0,\qquad \tau _{s}=C_{s}R_{s},
\label{SepSomaBoundary}
\end{equation}%
results in a transcendental equation%
\begin{equation}
\left[ 1-\frac{\tau _{s}}{\tau _{m}}\left( 1+\frac{R_{m}}{2R_{i}}\omega
^{2}\right) \right] \frac{R_{i}}{R_{s}}=\frac{C^{\prime }\left( x_{0},\omega
\right) +\frac{1}{2}\omega ^{2}S^{\prime }\left( x_{0},\omega \right) }{%
C\left( x_{0},\omega \right) +\frac{1}{2}\omega ^{2}S\left( x_{0},\omega
\right) }  \label{eigenvaluesomega}
\end{equation}%
for the eigenvalues $\omega .$ (There are infinitely many discrete
eigenvalues \cite{Churchill42} and \cite{Hartman73}, Ince?.) The
corresponding eigenfunctions $U_{n}=U\left( x,\omega _{n}\right)
=A_{n}u_{n}\left( x\right) $ are orthogonal%
\begin{equation}
\left( u_{m},u_{n}\right) =\delta _{mn}\left( u_{n},u_{n}\right)
\label{orthogonal}
\end{equation}%
with respect to an inner product that is given in terms of the
Lebesgue--Stieltjes integral \cite{Churchill42} (see also Appendix and \cite%
{Kellogg21}, \cite{Reid32} and \cite{Reid61}):%
\begin{eqnarray}
\left( u,v\right) &:&=\int_{x_{0}}^{x_{1}}u\left( x\right) v\left( x\right)
\ r\left( x\right) ds  \label{InnerProduct} \\
&&+\frac{1}{2}r^{2}\left( x_{1}\right) u\left( x_{1}\right) v\left(
x_{1}\right) +\frac{C_{s}}{2C_{m}}r^{2}\left( x_{0}\right) u\left(
x_{0}\right) v\left( x_{0}\right) .  \notag
\end{eqnarray}

A formal solution of the corresponding initial value problem takes the form%
\begin{eqnarray}
V\left( x,t\right) &=&V\left( x,\infty \right)  \label{GenSolIVP} \\
&&+\sum_{n}A_{n}\exp \left[ -\left( 1+\frac{R_{m}}{2R_{i}}\omega
_{n}^{2}\right) \frac{t}{\tau _{m}}\right] u_{n}\left( x\right) ,  \notag
\end{eqnarray}%
where $V\left( x,\infty \right) $ is the steady-state solution, $\omega
=\omega _{n}$ are roots of the transcendental equation (\ref%
{eigenvaluesomega}) and the corresponding eigenfunctions are given by%
\begin{equation}
u_{n}\left( x\right) =C\left( x,\omega _{n}\right) +\frac{1}{2}\omega
_{n}^{2}S\left( x,\omega _{n}\right) .  \label{eigenfunction}
\end{equation}%
Coefficients $A_{n}$ can be obtained by methods of Refs.~\cite{Churchill42}
and \cite{Durand84} with the help of the modified orthogonality relation (%
\ref{orthogonal}) as follows%
\begin{equation}
A_{n}=\frac{\left( V\left( x,0\right) -V\left( x,\infty \right) ,u_{n}\left(
x\right) \right) }{\left( u_{n}\left( x\right) ,u_{n}\left( x\right) \right)
}.  \label{An}
\end{equation}%
Substitution of (\ref{An}) into (\ref{GenSolIVP}) and changing the order of
summation and integration result in%
\begin{eqnarray}
V\left( x,t\right) &=&V\left( x,\infty \right)  \label{IntegralSol} \\
&&+\int_{\text{Supp\ }\mu }G\left( x,y,t\right) \left( V\left( y,0\right)
-V\left( y,\infty \right) \right) \ d\mu \left( y\right) ,  \notag
\end{eqnarray}%
where%
\begin{equation}
G\left( x,y,t\right) =\sum_{n}\exp \left[ -\left( 1+\frac{R_{m}}{2R_{i}}%
\omega _{n}^{2}\right) \frac{t}{\tau _{m}}\right] \frac{u_{n}\left( x\right)
u_{n}\left( y\right) }{\left\Vert u_{n}\right\Vert ^{2}}  \label{GreanFunc}
\end{equation}%
is an analog of the heat kernel. Infinite speed of propagation. Method of
images.

\subsection{A Single Branch with Piecewise Typering}

In the case of a piecewise tapering%
\begin{equation}
r=\left\{
\begin{array}{cc}
r_{0}\left( x\right) , & x_{0}\leq x\leq x_{01} \\
r_{1}\left( x\right) , & x_{01}\leq x\leq x_{1}%
\end{array}%
\right.  \label{2peace}
\end{equation}%
with $r_{0}\left( x_{01}^{-}\right) =r_{1}\left( x_{01}^{+}\right) ,$ in a
similar fashion, one can write%
\begin{equation}
U\left( x,\omega \right) =\left\{
\begin{array}{cc}
A_{0}\left[ C_{0}\left( x,\omega \right) -\left( \frac{R_{i}}{R_{s}}\left(
\frac{\tau _{s}}{\tau _{m}}-1\right) +\frac{C_{s}}{2C_{m}}\omega ^{2}\right)
S_{0}\left( x,\omega \right) \right] , & x_{0}\leq x\leq x_{01} \\
A_{1}\left[ C_{1}\left( x,\omega \right) +\frac{1}{2}\omega ^{2}S_{1}\left(
x,\omega \right) \right] , & x_{01}\leq x\leq x_{1}%
\end{array}%
\right.  \label{U2peace}
\end{equation}%
provided $C_{0}\left( x_{0},\omega \right) =1,$ $C_{0}^{\prime }\left(
x_{0},\omega \right) =0$ and $S_{0}\left( x_{0},\omega \right) =0,$ $%
S_{0}^{\prime }\left( x_{0},\omega \right) =1$ and $C_{1}\left( x_{1},\omega
\right) =1,$ $C_{1}^{\prime }\left( x_{1},\omega \right) =0$ and $%
S_{1}\left( x_{1},\omega \right) =0,$ $S_{1}^{\prime }\left( x_{1},\omega
\right) =1.$ Continuity and smoothness of the solution at the point $x_{01},$
namely,%
\begin{equation}
\frac{U^{\prime }\left( x_{01}^{-}\right) }{U\left( x_{01}^{-}\right) }=%
\frac{U^{\prime }\left( x_{01}^{+}\right) }{U\left( x_{01}^{+}\right) },
\label{cont}
\end{equation}%
results in the following equation for the eigenvalues%
\begin{equation}
\frac{C_{0}^{\prime }\left( x_{01},\omega \right) -\left( \frac{R_{i}}{R_{s}}%
\left( \frac{\tau _{s}}{\tau _{m}}-1\right) +\frac{C_{s}}{2C_{m}}\omega
^{2}\right) S_{0}^{\prime }\left( x_{01},\omega \right) }{C_{0}\left(
x_{01},\omega \right) -\left( \frac{R_{i}}{R_{s}}\left( \frac{\tau _{s}}{%
\tau _{m}}-1\right) +\frac{C_{s}}{2C_{m}}\omega ^{2}\right) S_{0}\left(
x_{01},\omega \right) }=\frac{C_{1}^{\prime }\left( x_{01},\omega \right) +%
\frac{1}{2}\omega ^{2}S_{1}^{\prime }\left( x_{01},\omega \right) }{%
C_{1}\left( x_{01},\omega \right) +\frac{1}{2}\omega ^{2}S_{1}\left(
x_{01},\omega \right) }.  \label{Omega2piece}
\end{equation}%
Introducing%
\begin{equation}
u_{n}\left( x\right) =\left\{
\begin{array}{cc}
u_{n}^{\left( 0\right) }\left( x\right) /u_{n}^{\left( 0\right) }\left(
x_{01}\right) , & x_{0}\leq x\leq x_{01} \\
u_{n}^{\left( 1\right) }\left( x\right) /u_{n}^{\left( 1\right) }\left(
x_{01}\right) , & x_{01}\leq x\leq x_{1}%
\end{array}%
\right. ,  \label{unpice}
\end{equation}%
where%
\begin{eqnarray*}
u_{n}^{\left( 0\right) }\left( x\right) &=&C_{0}\left( x,\omega _{n}\right)
-\left( \frac{R_{i}}{R_{s}}\left( \frac{\tau _{s}}{\tau _{m}}-1\right) +%
\frac{C_{s}}{2C_{m}}\omega _{n}^{2}\right) S_{0}\left( x,\omega _{n}\right) ,
\\
u_{n}^{\left( 1\right) }\left( x\right) &=&C_{1}\left( x,\omega _{n}\right) +%
\frac{1}{2}\omega ^{2}S_{1}\left( x,\omega _{n}\right) ,
\end{eqnarray*}%
one can obtain a formal solution in the form (\ref{IntegralSol})--(\ref%
{GreanFunc}) once again. Further details are left to the reader.

%Branching [Add and discuss branching rules at nodes. Quasi-steady solutions.]

\section{Summary}

In this Letter, we propose a simple graphical approach to steady state
solutions of the cable equation for a general model of dendritic tree with
tapering. A simple case of transient solutions is also briefly discussed.

\noindent \textbf{Acknowledgments.\/} We thank Carlos Castillo-Ch\'{a}vez,
Steve Baer, Hank Kuiper and Hal Smith for support, valuable discussions and
encouragement. This paper is written as a part of the summer 2010 program on
analysis of Mathematical and Theoretical Biology Institute (MTBI) and
Mathematical, Computational and Modeling Sciences Center (MCMSC) at Arizona
State University. The MTBI/SUMS Summer Research Program is supported by The
National Science Foundation (DMS-0502349), The National Security Agency
(DOD-H982300710096), The Sloan Foundation and Arizona State University.

\appendix

\section{Modified Orthogonality Relation}

We consider the Sturm--Liouville type problem,%
\begin{equation}
Lu+\lambda \rho u=0,  \label{A1}
\end{equation}%
for the second order differential operator%
\begin{equation}
Lu=\frac{d}{dx}\left[ k\left( x\right) \frac{du}{dx}\right] -q\left(
x\right) u,  \label{A2}
\end{equation}%
where $k,$ $q$ and $\rho $ are continuous real-valued functions on an
interval $\left[ x_{0},x_{1}\right] ,$ $k$ and $\rho $ are positive in $%
\left[ x_{0},x_{1}\right] ,$ $k^{\prime }$ exists and is continuous in $%
\left[ x_{0},x_{1}\right] ,$ subject to modified boundary conditions%
\begin{eqnarray}
u^{\prime }\left( x_{0}\right) +\left( a_{0}+b_{0}\lambda \right) u\left(
x_{0}\right)  &=&0,  \label{A4} \\
u^{\prime }\left( x_{1}\right) +\left( a_{1}-b_{1}\lambda \right) u\left(
x_{1}\right)  &=&0,  \notag
\end{eqnarray}%
where $a_{0},b_{0}\geq 0$ and $a_{1},b_{1}\geq 0$ are constants. With the
help of the second Green's formula (see, for example, \cite{NiPDE}),%
\begin{equation}
\int_{x_{0}}^{x_{1}}\left( vLu-uLv\right) \ dx=\left. k\left( v\frac{du}{dx}%
-u\frac{dv}{dx}\right) \right\vert _{x_{0}}^{x_{1}},  \label{A3}
\end{equation}%
for two eigenfunctions $u$ and $v$ corresponding to different eigenvalues%
\begin{equation}
Lu+\lambda \rho u=0,\qquad Lv+\mu \rho v=0,\quad \lambda \neq \mu
\label{A4a}
\end{equation}%
one gets the following orthogonality relation \cite{Churchill42}:%
\begin{equation}
\int_{x_{0}}^{x_{1}}u\left( x\right) v\left( x\right) \ \rho dx+b_{1}k\left(
x_{1}\right) u\left( x_{1}\right) v\left( x_{1}\right) +b_{0}k\left(
x_{0}\right) u\left( x_{0}\right) v\left( x_{0}\right) =0.  \label{A5}
\end{equation}%
Here, the modified inner product%
\begin{eqnarray}
\left( u,v\right)  &:&=\int_{\text{Supp\ }\mu }uv\ d\mu   \label{A6} \\
&=&\int_{x_{0}}^{x_{1}}u\left( x\right) v\left( x\right) \ \rho
dx+b_{1}k\left( x_{1}\right) u\left( x_{1}\right) v\left( x_{1}\right)
+b_{0}k\left( x_{0}\right) u\left( x_{0}\right) v\left( x_{0}\right)   \notag
\end{eqnarray}%
is defined in terms of the Lebesgue--Stieltjes integral \cite{Kolm:Fom}. The
modified orthogonality relation (\ref{A5}) holds also in the case of a
piecewice continuous derivative $k^{\prime }$ on the interval $\left[
x_{0},x_{1}\right] .$

The junction of three branches (see Figure~2) can be considered in a similar
fashion. Suppose that%
\begin{equation}
L_{i}u_{i}+\lambda \rho _{i}u_{i}=0,\qquad L_{i}u=\frac{d}{dx}\left[
k_{i}\left( x\right) \frac{du}{dx}\right] -q_{i}\left( x\right) u\label{A7}
\end{equation}%
with $k=0,$ $1,$ $2$ for three corresponding branches, respectively, and
boundary conditions are given by%
\begin{eqnarray}
u^{\prime }\left( x_{0}\right) +\left( a_{0}+b_{0}\lambda \right) u\left(
x_{0}\right)  &=&0,\label{A8} \\
u^{\prime }\left( x_{1}\right) +\left( a_{1}-b_{1}\lambda \right) u\left(
x_{1}\right)  &=&0,  \notag \\
u^{\prime }\left( x_{2}\right) +\left( a_{2}-b_{2}\lambda \right) u\left(
x_{2}\right)  &=&0  \notag
\end{eqnarray}%
at the terminal ends. Introducing integration over the whole tree $T$ by
additivity,%
\begin{eqnarray}
&&\int_{T}\left( vLu-uLv\right) \ dx=\int_{x_{0}}^{x_{1}}\left(
v_{0}L_{0}u_{0}-u_{0}L_{0}v_{0}\right) \ dx\label{A9} \\
&&\qquad +\int_{x_{12}}^{x_{1}}\left( v_{1}L_{1}u_{1}-u_{1}L_{1}v_{1}\right)
\ dx+\int_{x_{12}}^{x_{2}}\left( v_{2}L_{2}u_{2}-u_{2}L_{2}v_{2}\right) \ dx,
\notag
\end{eqnarray}%
and applying the Green formula (\ref{A3}) for each branch, one gets%
\begin{eqnarray}
\int_{T}\left( vLu-uLv\right) \ dx &=&k_{0}\left( x_{12}\right) \left(
v_{0}\left( x_{12}\right) u_{0}^{\prime }\left( x_{12}\right) -u_{0}\left(
x_{12}\right) v_{0}^{\prime }\left( x_{12}\right) \right) \label{A9a} \\
&&-k_{1}\left( x_{12}\right) \left( v_{1}\left( x_{12}\right) u_{1}^{\prime
}\left( x_{12}\right) -u_{1}\left( x_{12}\right) v_{1}^{\prime }\left(
x_{12}\right) \right)   \notag \\
&&-k_{2}\left( x_{12}\right) \left( v_{2}\left( x_{12}\right) u_{2}^{\prime
}\left( x_{12}\right) -u_{2}\left( x_{12}\right) v_{2}^{\prime }\left(
x_{12}\right) \right)   \notag \\
&&-k_{0}\left( x_{0}\right) \left( v_{0}\left( x_{0}\right) u_{0}^{\prime
}\left( x_{0}\right) -u_{0}\left( x_{0}\right) v_{0}^{\prime }\left(
x_{0}\right) \right)   \notag \\
&&+k_{1}\left( x_{1}\right) \left( v_{1}\left( x_{1}\right) u_{1}^{\prime
}\left( x_{1}\right) -u_{1}\left( x_{1}\right) v_{1}^{\prime }\left(
x_{1}\right) \right)   \notag \\
&&+k_{2}\left( x_{2}\right) \left( v_{2}\left( x_{2}\right) u_{2}^{\prime
}\left( x_{2}\right) -u_{2}\left( x_{2}\right) v_{2}^{\prime }\left(
x_{2}\right) \right) .  \notag
\end{eqnarray}%
We shall assume that the following continuity conditions:%
\begin{eqnarray}
&&u_{0}\left( x_{12}\right) =u_{1}\left( x_{12}\right) =u_{2}\left(
x_{12}\right) ,\label{A9b} \\
&&k_{0}\left( x_{12}\right) u_{0}^{\prime }\left( x_{12}\right) =k_{1}\left(
x_{12}\right) u_{1}^{\prime }\left( x_{12}\right) =k_{2}\left( x_{12}\right)
u_{2}^{\prime }\left( x_{12}\right)   \notag
\end{eqnarray}%
hold at the branching point $x_{12}.$ In view of of the boundary conditions (%
\ref{A8}), the modified orthogonality relation takes the form%
\begin{eqnarray}
&&\int_{x_{0}}^{x_{1}}u\left( x\right) v\left( x\right) \ \rho
dx+\int_{x_{0}}^{x_{1}}u\left( x\right) v\left( x\right) \ \rho
dx+\int_{x_{0}}^{x_{1}}u\left( x\right) v\left( x\right) \ \rho dx\label{A10}
\\
&&\qquad +b_{0}k\left( x_{0}\right) u\left( x_{0}\right) v\left(
x_{0}\right) +b_{1}k\left( x_{1}\right) u\left( x_{1}\right) v\left(
x_{1}\right) +b_{2}k\left( x_{2}\right) u\left( x_{2}\right) v\left(
x_{2}\right) =0.  \notag
\end{eqnarray}%
The case of junction of $\left( n+1\right) $-branches (see Figure~3)is similar.
In general, for an arbitrary tree, one may conclude that only the terminal ends
shall add additional mass points to the measure, if the corresponding boundary
and continuity conditions hold. Further details are left to the reader.

\end{document}